                        \newif\ifpaper \newif\ifPDF               
                        \newif\ifboyscout  \newif\ifarticle       
                        \boyscoutfalse 
                        \articlefalse 
                        \paperfalse\PDFtrue 

\documentclass[final,3p,times,sort&compress]{elsarticle}	

\usepackage{amssymb}
\usepackage{amsmath}
\usepackage{graphicx}
\usepackage[pdftex,colorlinks]{hyperref}
\usepackage[nodots]{numcompress}

\ifpaper 
       \newcommand{\color}[1]{}       
       \newcommand{\wwwcb}[1]{{\tt ChaosBook.org#1}}
       
\else 
		\newcommand{\wwwcb}[1]{       
                  {\tt \href{http://ChaosBook.org#1}
              {ChaosBook.org#1}}}

	   \hypersetup{
   pdfauthor=Stefan Froehlich and Predrag Cvitanovic,
   pdfkeywords=complex Lorenz flow,
   pdftitle=Reducing continuous symmetries
   	   }
\fi

\ifboyscout 
  \newcommand{\toCB}{\marginpar{\footnotesize 2CB}}  
  \newcommand{\PC}[1]{\\{\color{red} [{Predrag: #1}]}\\}
  
  \newcommand{\SF}[1]{$\footnotemark\footnotetext{Stefan: #1}$}

\else 
  \newcommand{\toCB}{}
  \newcommand{\PC}[1]{}
  
  \newcommand{\SF}[1]{}
  
\fi  

\newcommand{\refref} [1] {Ref.~\cite{#1}}

\newcommand{\refrefs}[1] {Refs.~\cite{#1}}

\newcommand{\reffig} [1] {Fig.~\ref{#1}}

\newcommand{\refsect}[1] {Section~\ref{#1}}

\newcommand{\refappe}[1] {\ref{#1}}

\newcommand{\rf}     [1] {~\cite{#1}}
\newcommand{\refeq}  [1] {(\ref{#1})}

\newcommand{\beq}{\begin{equation}}
\newcommand{\continue}{\nonumber \\ }

\newcommand{\nnu}{\nonumber}
\newcommand{\eeq}{\end{equation}}
\newcommand{\ee}[1] {\label{#1} \end{equation}}
\newcommand{\bea}{\begin{eqnarray}}

\newcommand{\eea}{\end{eqnarray}}
\newcommand{\barr}{\begin{array}}
\newcommand{\earr}{\end{array}}

\newcommand{\etc}{{etc.}}       
\newcommand{\ie}{{i.e.}}        

\newcommand\PoincSec{Poincar\'e section}

\newcommand{\braket}[2]
		   {\langle{#1}\vphantom{#2}|\vphantom{#1}{#2}\rangle}

\newcommand{\Sset}{Inflection hyperplane}
\newcommand{\sset}{inflection hyperplane} 
\newcommand{\sspSing}{\ensuremath{\ssp^*}} 	
\newcommand{\sspRSing}{\ensuremath{\sspRed^*}} 	
\newcommand{\template}{template} 
\newcommand{\angVel}{angular velocity}
\newcommand{\angVels}{angular velocities}

\newcommand{\EQV}[1]{\ensuremath{\mathrm{E}_{#1}}}
\newcommand{\REQV}[2]{\ensuremath{\mathrm{TW}_{#1#2}}} 

\newcommand{\vel}{\ensuremath{v}}   
\newcommand{\timeStep}{\ensuremath{{\delta \tau}}}  
\newcommand{\id}{{\ \hbox{{\rm 1}\kern-.6em\hbox{\rm 1}}}}
\newcommand{\dmn}{-dimensional}  

\newcommand{\On}[1]{\ensuremath{\textrm{O}(#1)}}
\newcommand{\SOn}[1]{\ensuremath{\textrm{SO}(#1)}}         

\newcommand{\pSRed}{\ensuremath{\bar{\cal M}}} 
\newcommand{\sspRed}{\ensuremath{\bar{\ssp}}}    
\newcommand{\velRed}{\ensuremath{\bar{\vel}}}    
\newcommand{\slicep}{\ensuremath{\bar{\ssp}'}}   
\newcommand{\sliceTan}[1]{\ensuremath{\bar{t}{}'_{#1}}}    
\newcommand{\groupTan}{\ensuremath{t}}    
\newcommand{\Group}{\ensuremath{G}}         
\newcommand{\Fix}[1]{\ensuremath{\mathrm{Fix}\left(#1\right)}}
\newcommand{\Lg}{\ensuremath{T}}   
\newcommand{\LieEl}{\ensuremath{g}}  
\newcommand{\gSpace}{\ensuremath{{\bf \theta}}}   

\newcommand{\statesp}{state space}

\newcommand{\reals}{\mathbb{R}}

\newcommand{\pS}{\ensuremath{{\cal M}}}          
\newcommand{\ssp}{\ensuremath{x}}                

\newcommand{\cLe}{complex Lorenz equations}
\newcommand{\cLf}{complex Lorenz flow}
\newcommand{\CLe}{Complex Lorenz equations}

\newcommand{\KS}{Kuramoto-Siva\-shin\-sky}

\newcommand{\pCf}{plane Couette flow}

\newcommand{\po}{periodic orbit}

\newcommand{\eqv}{equilib\-rium}

\newcommand{\eqva}{equilib\-ria}

\newcommand{\reqv}{rela\-ti\-ve equilib\-rium}

\newcommand{\reqva}{rela\-ti\-ve equilib\-ria}

\newcommand{\reducedsp}{reduced state space}

\newcommand{\fixedsp}{fixed-point subspace}

\newcommand{\slice}{slice}
\newcommand{\Slice}{Slice}
\newcommand{\mslices}{method of slices}

\newcommand{\mframes}{method of moving frames}
\newcommand{\Mframes}{Method of moving frames}

\journal{Communications in Nonlinear Science and Numerical Simulation}
\begin{document}
\title{Reduction of continuous symmetries of chaotic flows
       by the method of slices}
\author{Stefan Froehlich}
\author{Predrag Cvitanovi\'{c}\corref{cor1}}
\ead{predrag@gatech.edu}
\cortext[cor1]{Corresponding author}
\ead[url]{ChaosBook.org}

\address{Center for Nonlinear Science,
        School of Physics, Georgia Institute of Technology,
        Atlanta, GA 30332-0430}

\date{\today}

\begin{abstract}
  %
We study continuous symmetry reduction of dynamical systems
by the \mslices\ (\mframes) and show that a `slice'
defined by minimizing the distance to a single generic `{\template}'
intersects the group orbit of every point in the full {\statesp}. Global
symmetry reduction by a single slice is, however, not natural for a
chaotic / turbulent flow; it is better to cover the \reducedsp\ by a set
of slices, one for each dynamically prominent unstable pattern.
Judiciously chosen, such tessellation eliminates the singular traversals
of the \sset\ that comes along with each slice, an artifact of using the
{\template}'s local group linearization globally. We compute the jump in
the \reducedsp\ induced by crossing the \sset. As an illustration of the
method, we reduce the $\SOn{2}$ symmetry of the \cLe.
\end{abstract}

\begin{keyword}
	symmetry reduction; equivariant dynamics; relative equilibria;
	relative periodic orbits; slices; moving frames; Lie groups
\end{keyword}

\maketitle 

\section{Introduction}
    \label{sec:intro}

In spatially-extended turbulent flows one observes similar patterns at
different spatial positions and at different times. How `similar?' If the
flow is equivariant under a group of continuous symmetries, one way of
answering this question is by measuring distances between different
states in the symmetry-reduced \statesp\ $\pS/\Group$, a space in which
each group orbit (class of physically equivalent states) is represented
by a single point. This distance depends on the choice of norm and on
the symmetry-reduction method.

													\toCB
In 1980 Phil Morrison\rf{MorrGree80} showed how to derive Hamiltonian
description of ideal fluid (plasma) dynamics from the Low
Lagrangian\rf{Low58} by a Lie symmetry reduction, which in this context
amounts to the transformation from Lagrangian to Eulerian variables: the
\statesp\ of position-labeled Lagrangian trajectories of `fluid parcels'
is reduced to a much smaller \statesp\ of Eulerian velocity
fields. It is a
difficult example of reduction; the reduction steps have to be
executed judiciously, new variables cleverly chosen, and ``one should do
the Legendre transformations slowly and carefully when there are
degeneracies\rf{CHHM98}.'' Our goal here is different. Rather than to
reduce a particular set of dynamical equations, we seek to formulate a
computationally straightforward general method of reducing continuous
symmetries, applicable to any high-dimensional chaotic/turbulent flow,
such as the fluid flows bounded by pipes or planes. The
symmetry-reduction literature is very extensive (see
\refrefs{CBcontinuous,SiCvi10} for a review), but it basically offers two
approaches (a) invariant polynomial bases, and (b) methods which pick a
representative point by slicing group orbits, generalizing the way in
which {\PoincSec}s cut time-evolving trajectories. For high-dimensional
flows the \mslices\ studied in \refrefs{%
rowley_reduction_2003,%
CBcontinuous,SiminosThesis,SiCvi10,Wilczak09}
appears to be the only computationally feasible approach. Here
the method is rederived as a distance minimization problem in the space
of patterns.

The new results reported in this paper are:
    (a) A generic slice cuts across group orbits of {\em all}
        states in the \statesp\ (\refsect{sec:frame}).
    (b) Every slice carries along with it the {\sset}. We show how to
        compute the jump of the \reducedsp\ trajectory
         (\refsect{sec:mslices}) whenever it crosses
        through such singularity  (\refsect{sec:singul}).
    (c) We propose to avoid these singularities (artifacts of the symmetry
        reduction by the \mslices) by tiling the \statesp\ with an atlas
        constructed from a set of local slices  (\refsect{sec:chart}).
Pertinent facts about symmetries of dynamical systems are summarized in
\refappe{sec:SymmDyn}. In  \refappe{sec:singulProd} we show that for
continuous symmetries with product structure (such as $\SOn{2} \times
\SOn{2}$ symmetries of pipe and plane fluid flows), each symmetry induces
its own {\sset}.

In what follows we denote by `\mframes' the post-processing of the full
\statesp\ flow (\refsect{exam:CLErotAngle}), and by `\mslices' the
integration of flow confined to the \reducedsp\ (\refsect{sec:mslices}).
In practice, symmetry reduction is best carried out as post-processing,
after the numerical trajectory is obtained by integrating the full
\statesp\ flow. In particular, the symmetry-reduction induced
singularities (\refsect{sec:singul}) are more tractable numerically
if given the full \statesp\ trajectory.

 \begin{figure}
 \begin{center}
(a)\includegraphics[width=0.33\textwidth]{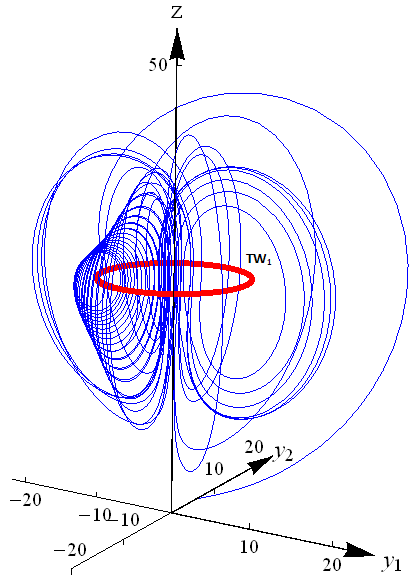}%
~~~~~~~~
(b)\includegraphics[width=0.32\textwidth]{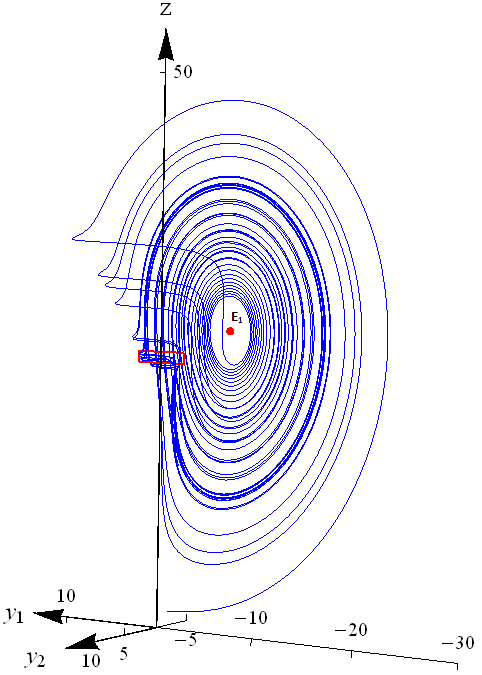}%
 \end{center}
 \caption{\label{fig:Fullspace}
(a) \CLe\ \refeq{eq:CLeR} exhibit a strange attractor for parameter
values \refeq{SiminosPrmts}, here projected on the $\{y_1,y_2,z\}$ axes.
{(thin line)} 
A segment of generic finite time trajectory.
{(thick line)} 
$\REQV{}{1}$, the only \reqv.
%
(b) The same strange attractor plotted in the symmetry-\reducedsp\
slice \refeq{PCsectQ}, defined by the group tangent $\sliceTan{}$ whose
choice is explained in \refeq{exmplTempl}. In the \reducedsp\ \reqv\
$\REQV{}{1}$ is reduced to \eqv\ $\EQV{1}$. Note, however, the
semicircular jumps in the reduced flow. These are analyzed in
\refsect{sec:singul}. For a blow-up of the jump indicted by the small
rectangle (red), see \reffig{fig:singpass}\,(a).
 }%
 \end{figure}

We shall illustrate symmetry reduction by applying it to the
5-dimensional \cLe\rf{GibMcCLE82}
\bea
	\dot{x}_1 &=& -\sigma x_1 + \sigma y_1
\,,\qquad\qquad\qquad
	\dot{x}_2 \,=\, -\sigma x_2 + \sigma y_2
\continue
	\dot{y}_1 &=& (r_1-z)\, x_1  - ~y_1 - e y_2
\,,\qquad\;
	\dot{y}_2 \,=\, (r_1-z)\, x_2 + e y_1 - ~y_2
\label{eq:CLeR}\\
	\dot{z}~ &=& -b z + x_1 y_1 + x_2 y_2
\,.
\nnu
\eea
In all numerical calculations that follow we shall set the
parameters to \refref{SiCvi10} values,
\beq
r_1=28,\; b={8}/{3},\;
\sigma=10,\quad \mbox{and}  \quad e={1}/{10}
\,,
\ee{SiminosPrmts}
for which the flow exhibits a strange attractor,
\reffig{fig:Fullspace}\,(a).
Our goal is to understand in detail this flow in the symmetry-\reducedsp,
\reffig{fig:Fullspace}\,(b), in particular the singularities induced by
the symmetry reduction.

    {
A flow $\dot{x}= \vel(x)$ is \emph{equivariant} under a coordinate
transformation $\LieEl$ if the form of the equations of motion is
preserved by the transformation,
\beq
\vel(x)=\LieEl^{-1}\vel(\LieEl \, x)
\,.
\ee{eq:FiniteRot1}
The totality of elements
$\LieEl$ forms \Group, the {\em symmetry group} of the flow.
    }
The \cLe\ are a simple example of a dynamical
system with a continuous (but no discrete) symmetry. They are equivariant
\refeq{eq:FiniteRot1} under \SOn{2} rotations by
	\ifarticle  
\bea
\LieEl(\gSpace)
    &=&
\exp{({\gSpace} \cdot \Lg)}
	 \,=\,
  \left(\barr{ccccc}
  \cos \gSpace  & \sin \gSpace  & 0 & 0 & 0 \\
 -\sin \gSpace  & \cos \gSpace  & 0 & 0 & 0 \\
 0 & 0 &  \cos \gSpace & \sin \gSpace   & 0 \\
 0 & 0 & -\sin \gSpace & \cos \gSpace   & 0 \\
 0 & 0 & 0             & 0              & 1
    \earr\right)
\continue
\Lg &=&
   \left(\barr{ccccc}
    0  &  1 & 0  &  0 & 0  \\
   -1  &  0 & 0  &  0 & 0 \\
    0  &  0 & 0  &  1 & 0  \\
    0  &  0 &-1  &  0 & 0 \\
    0  &  0 & 0  &  0 & 0
    \earr\right)
\label{CLfRots}
\eea
    \else  
\bea
\LieEl(\gSpace)
    &=&
\exp{({\gSpace} \cdot \Lg)}
	 \,=\,
  \left(\barr{ccccc}
  \cos \gSpace  & \sin \gSpace  & 0 & 0 & 0 \\
 -\sin \gSpace  & \cos \gSpace  & 0 & 0 & 0 \\
 0 & 0 &  \cos \gSpace & \sin \gSpace   & 0 \\
 0 & 0 & -\sin \gSpace & \cos \gSpace   & 0 \\
 0 & 0 & 0             & 0              & 1
    \earr\right)
\,,\qquad \Lg \,=\,
   \left(\barr{ccccc}
    0  &  1 & 0  &  0 & 0  \\
   -1  &  0 & 0  &  0 & 0 \\
    0  &  0 & 0  &  1 & 0  \\
    0  &  0 &-1  &  0 & 0 \\
    0  &  0 & 0  &  0 & 0
    \earr\right)
\label{CLfRots}
\eea
	\fi
(for group-theoretical notation, see \refappe{sec:SymmDyn}). The group is
1\dmn\ and compact, its elements parameterized by $\gSpace \mbox{ mod }
2\pi$. The fixed-point subspace \refeq{dscr:InvPoints} is the $z$-axis.
The velocity \refeq{eq:CLeR} at a point on the $z$-axis points only in
the $z$-direction and so the trajectory remains on the $z$-axis for all
times. The action of \SOn{2}\ thus decomposes the  \statesp\ into $m=0$
invariant subspace ($z$-axis) and  $m=1$ subspace of multiplicity
2. Locally, at \statesp\ point $\ssp$, the infinitesimal action of the
group is given by the group tangent field $\groupTan(\ssp) = \Lg \ssp =
(x_2,-x_1,y_2,-y_1,0)$, with the flow induced by the group action normal
to the radial direction in the $(x_1,x_2)$ and $(y_1,y_2)$ planes, while
the $z$-axis is left invariant.

\section{\Mframes}
    \label{sec:frame}

Suppose you are observing turbulence in a pipe flow, or your
defibrillator has a mesh of sensors measuring electrical currents that
cross your heart, or you have a precomputed pattern, and are sifting
through the data set of observed patterns for something like it. Here you
see a pattern, and there you see a pattern that seems much like the first
one. How `much like the first one?' Think of the first pattern
(represented by a point {\slicep} in the \statesp\  \pS) as a
`template'%
\rf{rowley_reconstruction_2000,%
rowley_reduction_2003,%
ahuja_template-based_2007}
or a
`reference state' and use the symmetries of the flow to slide and rotate
the `{\template}' until it overlays the second pattern (a point $\ssp$ in
the \statesp), \ie, act with elements of the symmetry group \Group\ on
the {\template} $\slicep \to \LieEl(\gSpace)\,\slicep$ until the
distance between the two patterns
\beq
|\ssp - \LieEl(\gSpace)\,\slicep|
    = |\sspRed - \slicep|
\label{minDistance}
\eeq
is minimized. Here $\sspRed$ is the point on the group orbit of $\ssp$
(the set of all points that $\ssp$ is mapped to under the group
actions),
\beq
\ssp=\LieEl(\gSpace)\,\sspRed
	\,,\qquad
\LieEl \in \Group
\,,
\ee{sspOrbit}
closest to the {\template} {\slicep}, the Lie group element
$\LieEl=\LieEl(\gSpace)\propto\exp{({\gSpace} \cdot \Lg)}$ is
parameterized by angles $\gSpace =
(\gSpace_1,\gSpace_2,\cdots\gSpace_N)$, and the distance is an invariant
of the symmetry group, $|\LieEl\ssp|^2=|\ssp|^2$. We assume that \Group\
is a subgroup of the group of orthogonal transformations
$\On{d}$, and measure
distance $|\ssp|^2=\braket{\ssp}{\ssp}$ in terms of the Euclidean inner
product
\( 
\braket{x}{y} = \sum_i^d {x}_i y_i
	\,.
\) 
Its Lie algebra {generators} $\Lg_a$ \refeq{FiniteRot} are $N$
linearly independent $[d\!\times\!d]$ antisymmetric matrices acting
linearly on the {\statesp} vectors $\ssp \in \pS \subset \reals^d$.

If the \statesp\ is a normed function space,
\( 
\braket{h}{f} = \int dx \, h(x) f(x)
\,,
\) 
one customarily measures distance between two patterns in the $L^2$ norm,
$|f|^2 = \braket{f}{f}$. In computations, spatially-extended functions are
represented by discrete meshes or finite basis sets, within a (possibly
large) finite-dimensional \statesp\  $\pS \subset \reals^d$. An example
is representation of a dissipative PDE by truncating the Fourier basis
\refeq{FourierExp} to a finite number of modes.

\begin{figure}
 \begin{center}
  \setlength{\unitlength}{0.40\textwidth}
(a)
  \begin{picture}(1,0.87085079)%
    \put(0,0){\includegraphics[width=\unitlength]{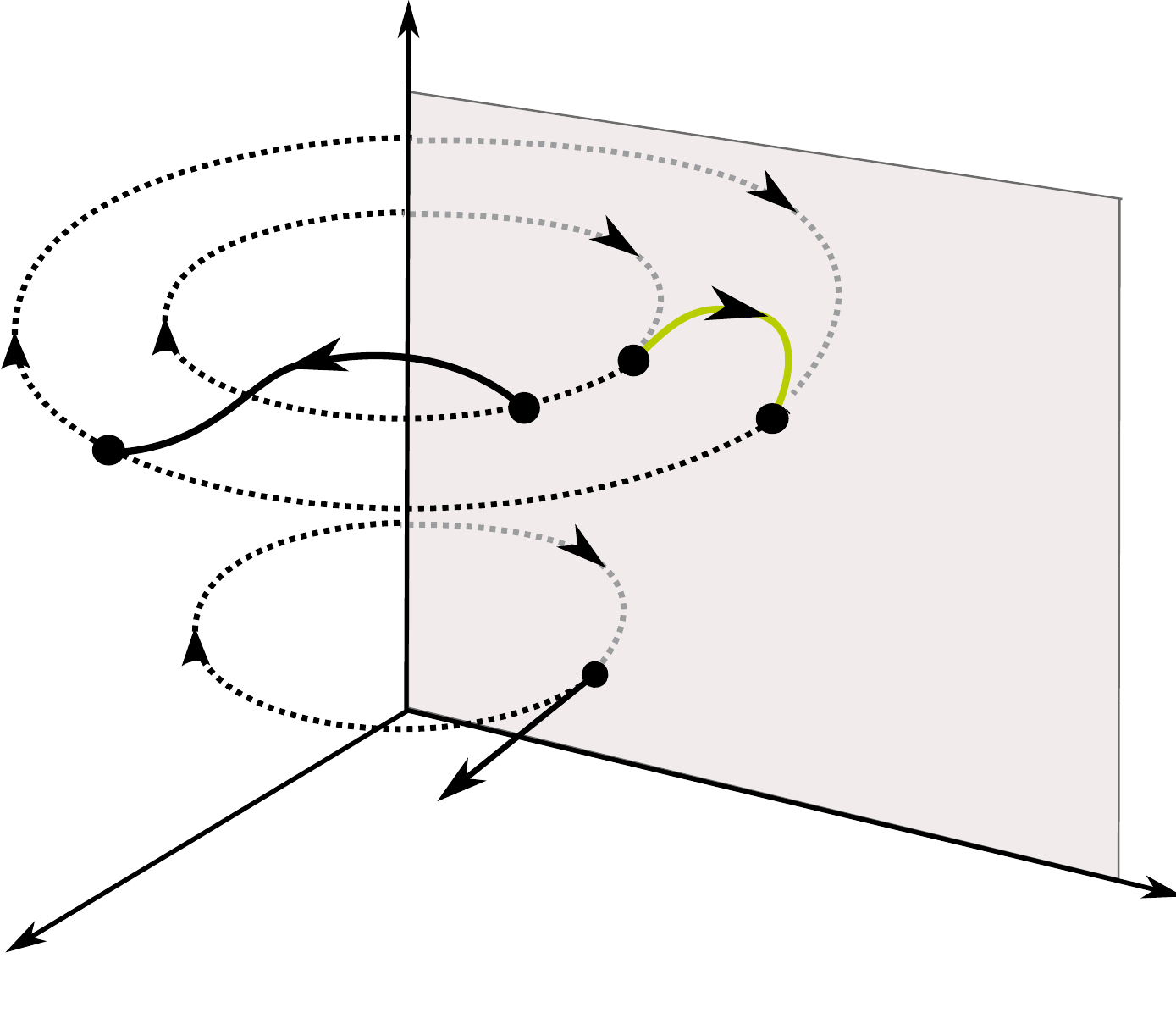}}%
    \put(0.82835155,0.19007659){\color[rgb]{0,0,0}\rotatebox{-14.84025424}{\makebox(0,0)[lb]{\smash{$\pSRed$}}}}%
    \put(0.07077338,0.28688228){\color[rgb]{0,0,0}\rotatebox{0.0313674}{\makebox(0,0)[lb]{\smash{$\LieEl\,\slicep$}}}}%
    \put(0.53023327,0.26593335){\color[rgb]{0,0,0}\rotatebox{0.0313674}{\makebox(0,0)[lb]{\smash{$\slicep$}}}}%
    \put(0.4284954,0.179285){\color[rgb]{0,0,0}\rotatebox{0.0313674}{\makebox(0,0)[lb]{\smash{$\sliceTan{}$}}}}%
    \put(0.00798985,0.42305068){\color[rgb]{0,0,0}\rotatebox{0.0313674}{\makebox(0,0)[lb]{\smash{$\ssp(\tau)$}}}}%
    \put(0.65766235,0.45412105){\color[rgb]{0,0,0}\rotatebox{0.0313674}{\makebox(0,0)[lb]{\smash{$\sspRed(\tau)$}}}}%
    \put(0.06916446,0.74280851){\color[rgb]{0,0,0}\rotatebox{0.0313674}{\makebox(0,0)[lb]{\smash{$\LieEl(\tau)$}}}}%
  \end{picture}%
~~~
(b)
  \begin{picture}(1,0.8708158)%
    \put(0,0){\includegraphics[width=\unitlength]{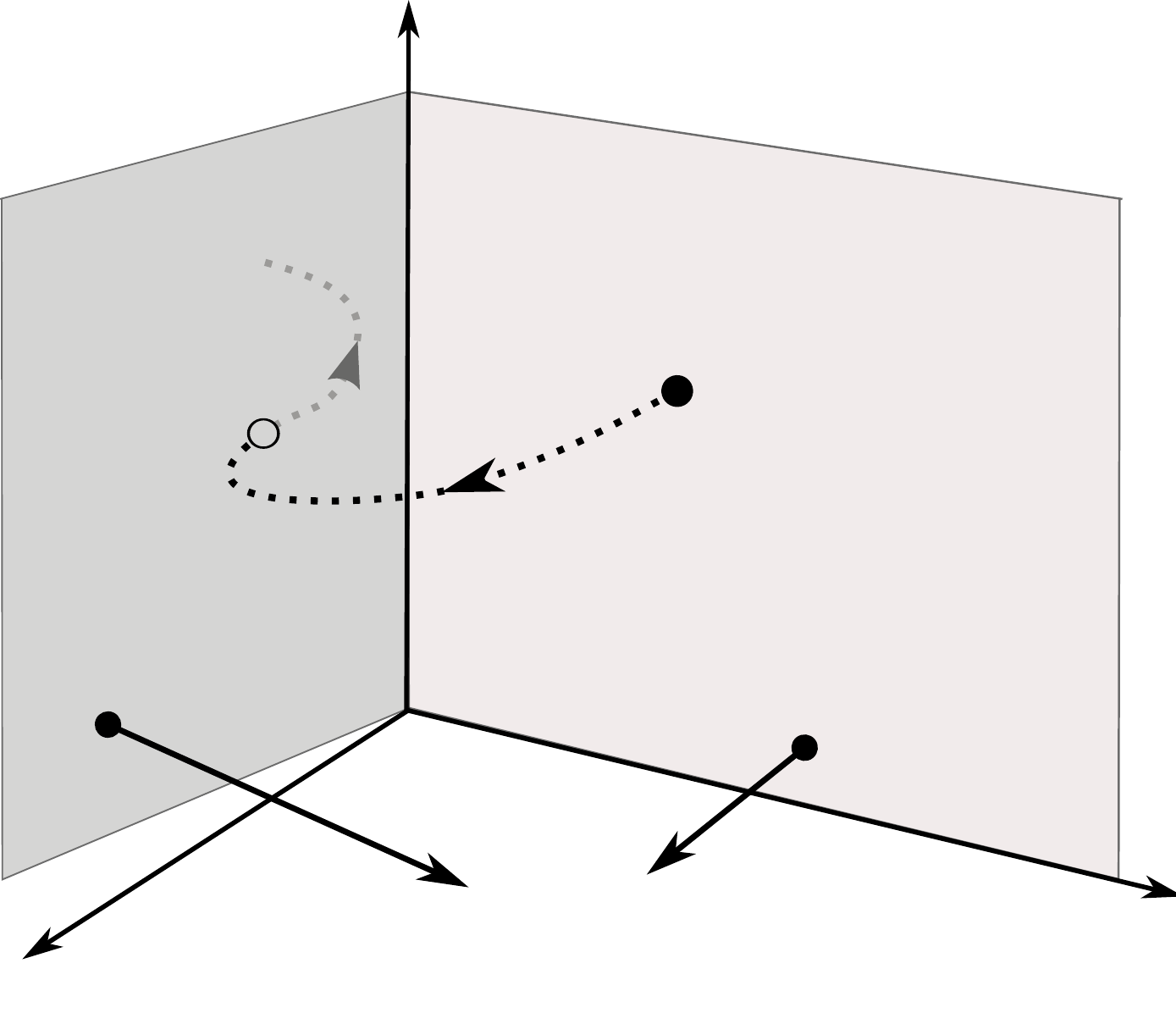}}%
    \put(0.7987485,0.62071555){\color[rgb]{0,0,0}\rotatebox{-9.7016004}{\makebox(0,0)[lb]{\smash{$\pSRed$}}}}%
    \put(0.38209831,0.39016168){\color[rgb]{0,0,0}\rotatebox{0.0313674}{\makebox(0,0)[lb]{\smash{$\ssp = \LieEl\,\sspRed$}}}}%
    \put(0.14952406,0.52632455){\color[rgb]{0,0,0}\rotatebox{0.0313674}{\makebox(0,0)[lb]{\smash{$\sspSing$}}}}%
    \put(0.29583374,0.67396856){\color[rgb]{0,0,0}\rotatebox{18.56777832}{\makebox(0,0)[lb]{\smash{$S$}}}}%
    \put(0.21967016,0.09601742){\color[rgb]{0,0,0}\rotatebox{0.0313674}{\makebox(0,0)[lb]{\smash{$\Lg^2\slicep$}}}}%
    \put(0.70853504,0.20366515){\color[rgb]{0,0,0}\rotatebox{0.0313674}{\makebox(0,0)[lb]{\smash{$\slicep$}}}}%
    \put(0.60680126,0.11702028){\color[rgb]{0,0,0}\rotatebox{0.0313674}{\makebox(0,0)[lb]{\smash{$\sliceTan{}$}}}}%
    \put(0.60809768,0.50392771){\color[rgb]{0,0,0}\rotatebox{0.0313674}{\makebox(0,0)[lb]{\smash{$\sspRed$}}}}%
  \end{picture}%
 \end{center}
 \caption{\label{fig:slice}
The \mframes.
(a)
\Slice\ {$\pSRed$ $(\,\,\supset \pS/\Group)$}
{lies in the $(d\!-\!N)$\dmn\ half-}hyperplane
\refeq{PCsectQ} normal to $\sliceTan{}$, where $\sliceTan{}$ is
the  $N$\dmn\ tangent to the group orbit $\LieEl\,\slicep$ (dotted line) of the
{\template} point  $\slicep$, evaluated at $\slicep$. This is a highly idealized
sketch: A group orbit is an $N$\dmn\ manifold, and even for $\SOn{2}$ it
is usually only topologically a circle, and can intersect a {hyperplane}
any number of times. {Such hyperplane} intersects {\em all} full
\statesp\ group orbits (indicated by dotted lines here).  The full
\statesp\ trajectory $\ssp(\tau)$ and the \reducedsp\ trajectory
$\sspRed(\tau)$ are equivalent up to a `moving frame' rotation
$\ssp(\tau)=\LieEl(\tau)\,\sspRed(\tau)$, where $\LieEl(\tau)$ is
a shorthand for $\LieEl(\gSpace(\tau))$.
(b)
For $\SOn{2}$ two hyperplanes are associated with  a given {\template}
\slicep; the slice $\pSRed$, and the hyperplane of points $\sspSing$
normal to the quadratic Casimir-weighted vector $\Lg^2\slicep$, for which
the curvature \refeq{SO2inflPoint} of the distance function
\refeq{minDistance} changes sign. This defines a group-theoretic
boundary of the {\template} neighborhood: For rotation angles $\gSpace$
beyond this boundary the group orbit $\LieEl(\gSpace)\,\ssp$ has left the
neighborhood. The intersection of the two hyperplanes is the {\em \sset}
$\sspRSing \in S$, within which all group tangents $\groupTan(\sspRSing)$
point into the slice and are thus normal to $\sliceTan{}$. For the
lack of dimensions, the intersection $S$ is drawn here as a `line,'
the $z$ axis in this 3\dmn\ sketch. $S$ is actually
a $(d\!-\!2)$\dmn\ hyperplane, but that is not easy to visualize.
 }%
 \end{figure}

The minimal distance is a solution of the extremum conditions
\[ 
\frac{\partial ~~}{\partial \gSpace_a} |\ssp - \LieEl(\gSpace)\,\slicep|^2
   =
    {
2\, \braket{\sspRed - \slicep}{\sliceTan{a}}
    }
   = 0
    \,,\qquad
	  \sliceTan{a} = \Lg_a \slicep
\,.
\] 
By the antisymmetry of the Lie algebra generators we have
$\braket{\slicep}{\sliceTan{a}} = \braket{\slicep}{\Lg_{a}\slicep}=0$, so
we can replace
    {
$\sspRed - \slicep \to \sspRed$,
    }
and the `moving frame' transformation
parameters $\gSpace$ which map the state $\ssp$ to $\sspRed$, the group
orbit point closest to the {\template} $\slicep$, satisfy
\beq
\braket{\sspRed}{\sliceTan{a}} =0
    \,,\qquad
\LieEl(\gSpace)\,\sspRed = \ssp
\,.
\ee{PCsectQ}
    {
Thus the set of \emph{extremal} group orbit points
        }
lies in a $(d\!-\!N)$\dmn\ hyperplane, the set of vectors
orthogonal to the {\template} tangent space spanned by tangent vectors
$\{\sliceTan{1},\cdots,\sliceTan{N}\}$
                    \toCB
\beq
\sspRed_1\sliceTan{a,1} + \sspRed_2\sliceTan{a,2}
  + \cdots + \sspRed_d\sliceTan{a,d} = 0
\,,
\ee{hyperpl}
													\toCB
see \reffig{fig:slice}\,(a).
    {
This hyperplane contains different types of extremal points. For example,
the point \emph{furthest} away from the \template\ \slicep\ also
satisfies the extremal conditions. While group orbits are embedded into
the high-dimensional full \statesp\ in a highly convoluted manner, this
hyperplane is a linear section through them, a global extension of the
tangent space of \slicep, which can be a good description of the
`similarity' to a \template\ only in a local neighborhood of \slicep. Our
goal is to reduce the symmetry of the flow by slicing the totality of
group orbits by a small set of such neighborhoods, one for each distinct
\template, with each group orbit sliced only once. Group orbits close to
\slicep\ cross the hyperplane transversely; the border of the
neighborhood is defined by group orbits that reach the hyperplane
tangentially. In case of a local Poincar\'e section, determination of
such border is a nontrivial task, but as we shall see in
\refsect{sec:singul}, for group orbits this border is easy to determine.

The set of the group orbit points \emph{closest} to the \template\ \slicep\
form an open connected neighborhood
of \slicep, a neighborhood in which each group orbit intersects the
hyperplane \emph{only once}.
As we shall show in \refsect{sec:singul}, this neighborhood is contained in
a half-hyperplane, bounded on one side by the intersection of \refeq{hyperpl}
with its {\sset}.
In what follows we shall refer to this connected open neighborhood
of \slicep\ as a \emph{\slice} $\pSRed_{\slicep} \supset \pS/\Group$,
and
to  \refeq{PCsectQ} as the \emph{slice conditions}.
        }
\Slice\ so defined is
a particular case of symmetry reduction by transverse sections of group
orbits\rf{FelsOlver98,FelsOlver99,OlverInv} that can be traced back to
Cartan's \mframes\rf{CartanMF}. \emph{Moving frame} refers to the action
$\LieEl(\gSpace)$ that brings a \statesp\ point \ssp\ into the slice.
We denote the full \statesp\ points and velocities by $\ssp$,
$\vel(\ssp)$, and the \reducedsp\  points and velocities by $\sspRed$,
$\velRed(\sspRed)$.

In the choice of the {\template} one should avoid solutions
that belong to the invariant or partially symmetric subspaces; for such
choices $\Lg_{a}\slicep=0$, and some or all $\sliceTan{a}$ vanish identically
and impose no slice conditions. The {\template} $\slicep$ should be a
generic \statesp\ point in the sense that its group orbit has the full
$N$ dimensions of the group \Group. In particular, even though the
simplest solutions (laminar, \etc) often capture important physical
features of a flow, most \eqva\ and short \po s have nontrivial
symmetries and thus are not suited as choices of symmetry-reducing
{\template s}.
It should also be emphasized that in general a {\template} is \emph{not}
a spatially-{localized} structure. We are not using translations /
rotations to superimpose a localized, `solitonic' solution over a
localized {\template}. In a strongly nonlinear, turbulent flow a good
{\template} is typically a nontrivial global solution.

In summary: given the minimum Euclidean distance condition, the
point $\sspRed$ in the group orbit of $\ssp$ closest
to the {\template} $\slicep$ lies in a \emph{slice}, a {\em hyperplane}
normal to the group action tangent space $\sliceTan{}$, for any \statesp\
point $\ssp \in \pS$. {\em Symmetry reduction} by the \mframes\ is a
precise rule for how to pick a unique point \sspRed\ for each such
symmetry equivalence class, and compute the \emph{moving frame}
transformation $\ssp =\LieEl(\gSpace)\, \sspRed$ that relates the full
\statesp\ point  $\ssp \in \pS$ to its symmetry reduced representative
$\sspRed \in \pSRed$.

\subsection{Computing the moving frame rotation angle}
\label{exam:CLErotAngle}

															\toCB
The idea of reducing a flow with Lie-group structure to a system of a
smaller dimension dates back to Sophus Lie.
Time-evolution and symmetry group actions foliate the \statesp\ into
$(N\!+\!1)$\dmn\ submanifolds: Given a state (a \statesp\ point $\ssp(0)$
at time $\tau=0$), we can trace its  1\dmn\ trajectory $\ssp(\tau)$ by
integrating its equations of motion, and its $N$\dmn\ group orbit by
acting on it with the symmetry group \Group. Locally, a continuous time
flow can be reduced by a \PoincSec; a slice does the same for local
neighborhoods of group orbits.

To show how the rotation into the \slice\ is computed, consider first the
\cLe. Substituting the \SOn{2}\ Lie algebra
generator and a finite angle \SOn{2} rotation \refeq{CLfRots} acting on a
5\dmn\ \statesp\ into the slice condition \refeq{PCsectQ}
yields
{
\(\braket{\ssp}{\sliceTan{}}\cos\gSpace
    -\braket{\groupTan_{}(\ssp)}{\sliceTan{}} \sin\gSpace
= 0
\,,
\)}
and the explicit formula for frame angle $\gSpace$:
\bea
\tan\gSpace &=&
   {\braket{\ssp}{\sliceTan{}}}/
          {\braket{\groupTan_{}(\ssp)}{\sliceTan{}}}
\,.
\label{SL:CLEsliceRot}
\eea
The dot product of two tangent fields in \refeq{SL:CLEsliceRot} is a
sum of inner products weighted by Casimirs \refeq{QuadCasimir},
    {
\beq
\braket{\groupTan(\ssp)}{\groupTan(\slicep)}
   = \sum_m C_2^{(m)} {\ssp}_i\, \delta_{ij}^{(m)} \slicep_j
\,.
\ee{braket}
    }
For the \cLe\
$\ssp = (x_1,x_2,y_1,y_2,z)$,
    {
$\slicep = (\bar{x}_1',\bar{x}_2',\bar{y}_1',\bar{y}_2',\bar{z}')$,
     }
and applying the moving frame condition \refeq{SL:CLEsliceRot} yields
    {
\beq
\tan\gSpace =
\frac{x_1 \bar{x}_2'-x_2 \bar{x}_1'+y_1 \bar{y}_2' -y_2 \bar{y}_1'}
       {x_1 \bar{x}'_1+x_2 \bar{x}'_2+y_1 \bar{y}'_1+y_2 \bar{y}'_2}
\,.
\ee{braketCL}
    }
This formula is particularly simple, as in the \cLe\
example the group acts only through $m=0$ and $m=1$ representations
(in the Fourier mode labeling of \refeq{SO2irrepAlg-Lg}).

Consider next the general form \refeq{SO2irrepAlg-m} of action of an
$\SOn{2}$ symmetry on arbitrary Fourier coefficients of a spatially
periodic function \refeq{FourierExp}. Substituting this into the slice
condition \refeq{PCsectQ}
{
and using $g^{(m)}(\gSpace)=\cos(m\gSpace)\id^{(m)} +\sin(m\gSpace)
\frac{1}{m}\Lg^{(m)}$, see \refeq{SO2irrepAlg-m}}, we find that
\bea
\braket{e^{-\gSpace \Lg}\ssp}{\groupTan(\slicep)}
=\braket{\ssp}{\sum\limits_m \left(\cos(m\gSpace) \id^{(m)}
     +\sin(m\gSpace) \frac{1}{m}\Lg^{(m)}\right) \sliceTan{}}
\continue
{=\sum\limits_m
   \left(
    \braket{\ssp}{\Lg^{(m)} \slicep} \cos(m\gSpace)
  - m\braket{\ssp}{\id^{(m)} \slicep} \sin(m\gSpace)
   \right)}
   =0
\,.
\label{eq:so2sing}
\eea
This is a polynomial equation, with coefficients determined by
$\braket{\ssp}{\Lg^{(m)} \slicep}$ and $\braket{\ssp}{\id^{(m)}\slicep}$,
as we can see by rewriting $\cos(m\gSpace)$, $\sin(m\gSpace)$ as
polynomials of degree $m$ in $\sin(\gSpace)$ and $\cos(\gSpace)$. Each
phase $\gSpace$ that rotates $\ssp$ into any of the group-orbit
traversals of the slice hyperplane corresponds to a real root of this
polynomial.

As a generic group orbit is a smooth $N$\dmn\ manifold embedded in the
$d$\dmn\ \statesp, several values of $\gSpace$ might be local extrema of
the distance function \refeq{minDistance}.
Our prescription is to pick the closest \reducedsp\ point as the unique
representative of the entire group orbit. \ie, determine the global
minimum (infimum) of distance \refeq{minDistance}.
For example, group orbits of
\SOn{2}\ are topologically circles, and the distance function
has maxima, minima and inflection points as {critical points}:
if \gSpace\ is a solution of the slice condition \refeq{SL:CLEsliceRot}
for \cLe,
so is $\gSpace+\pi$. We can pick the closest by noting that
the local minima have positive curvature,
\beq
\frac{\partial^2}
     {\partial \gSpace^2}
        |\sspRed - \slicep|^2
    =
{- 2 \, \braket{\sspRed}{\Lg^2\slicep}}
\,.
\ee{SO2inflPoint}
For the \cLe, this determines which moving frame angle will be used since
{the} distance function \refeq{minDistance} has
only a minimum and a maximum.
It does not matter
whether the group is compact, for example $\SOn{n}$, or noncompact, for
example the Euclidean group $E_2$ that underlies the generation of spiral
patterns\rf{Barkley94}; in either case any group orbit has one or several
locally closest passages to the {\template} state, and generically only
one that is the closest one.
(Here we focus only on continuous symmetries - discrete symmetries that
flows such as the \KS\ and {\pCf} exhibit will also have to be taken into
account\rf{SCD07,HGC08,DasBuchMirror}.)

However, `picking the closest' point in a group orbit of a pattern very unlike
the {\template} is not necessarily a sensible thing to do; as such state
evolves in time, distant points along its orbit can come closer to the
{\template}, causing discontinuous jumps in the moving frame angle. We
shall show in \refsect{sec:singul} that this is a generic phenomenon for a
single-slice symmetry reduction, and propose a cure in \refsect{sec:chart}.

In summary, we do not have to compute all zeros of the slice condition
\refeq{PCsectQ} - all we care about is the zero that
corresponds to the shortest distance \refeq{minDistance}.
While post-processing of a full \statesp\ trajectory $\ssp(\tau_j)$
requires a numerical (Newton method) determination of the
moving frame rotation
$\gSpace(\tau_j)$ at each time step $\tau_j$, the computation is not
as onerous as it might seem, as the knowledge of $\gSpace(\tau_j)$ and
$\groupTan(\sspRed(\tau_j))$
gives us a very good guess for $\gSpace(\tau_{j+1})$. We
can go a step further, and write the equations for the flow restricted to
the \reducedsp\ \pSRed.

\subsection{Dynamics within a slice}
\label{sec:mslices}

Any \statesp\ trajectory can be written in a factorized
form $\ssp(\tau)=\LieEl(\tau)\,\sspRed(\tau)$
(here $\LieEl(\tau)$ is a shorthand for $\LieEl(\gSpace(\tau))$,
or perhaps even $\LieEl(\gSpace(\ssp(\tau)))$).
Differentiating both sides with respect to time and
setting $\velRed={d\sspRed}/{d\tau}$ we find
\(
\vel(\ssp)=\dot{\LieEl} \, \sspRed+\LieEl \, \velRed(\sspRed)
\,.
\)
By the equivariance \refeq{eq:FiniteRot}
{\[
\vel(\sspRed)=\velRed(\sspRed) + \LieEl^{-1} \, \dot{\LieEl} \, \sspRed
\,.
\]}
Noting that $\LieEl^{-1}\dot{\LieEl}=e^{-\gSpace \cdot \Lg} \,
\frac{d ~~}{d \, \tau} e^{\gSpace \cdot \Lg}=\dot{\gSpace}\cdot \Lg$,
we obtain the equation for the velocity of the reduced flow:
\beq
\velRed(\sspRed)=\vel(\sspRed)-\dot{\gSpace}(\sspRed)\cdot \groupTan(\sspRed)
\,.
\ee{eq:redVel}
The velocity $\vel$ in the full \statesp\ is thus the sum of the
`angular' velocity \refeq{PC:groupTan1} along the group orbit,
    {
$\dot{\gSpace} \cdot \groupTan(\sspRed)$,
    }
and the remainder $\velRed$.

Eq. \refeq{eq:redVel} is true for any factorization
$\ssp=\LieEl \sspRed$, and by itself provides no
information on how to calculate $\dot{\gSpace}$. That is attained by
demanding that the reduced trajectory stays within a slice, by imposing
the slice conditions \refeq{PCsectQ}:
\beq
\braket{\vel(\sspRed)}{\sliceTan{a}}
 -\braket{\dot{\gSpace}\cdot \groupTan(\sspRed)}{\sliceTan{a}}=0
\,.
\label{eq:slicecondition}
\eeq
This is a matrix equation in
$\braket{\groupTan_b(\sspRed)}{\sliceTan{a}}$ that
    {
the authors of \refrefs{ahuja_template-based_2007,FiSaScWu96} claim one
can in principle solve for any Lie group. We consider here only the
$\SOn{2}$ case, which has a single group tangent:
        }
\bea
\velRed(\sspRed) &=& \vel(\sspRed)
   -\dot{\gSpace}(\sspRed) \, \groupTan(\sspRed)
\continue
\dot{\gSpace}(\sspRed) &=& {\braket{\vel(\sspRed)}{\sliceTan{}}}/
               {\braket{\groupTan(\sspRed)}{\sliceTan{}}}
\,.
\label{eq:so2reduced}
\eea
                                                    \toCB
One way to think about this reduction of a flow to a slice is in terms of
Lagrange multipliers (see {Stone and Goldbart}\rf{StGo09}, Sect 1.5 for
intuitive, geometrical interpretation of Lagrange multipliers). The first
equation defines the flow confined to the slice,
the `shape', `template' or `slice' dynamics\rf{rowley_reduction_2003}, (see
\reffig{fig:Fullspace}\,(b), \reffig{fig:slice}\,(a)),
and integration of the second,
`reconstruction' equation\rf{Marsd92,MarsdRat94} enables us to track the
corresponding trajectory in the full \statesp. For invariant subspaces
$\dot{\gSpace}=0$, so they are always included within the slice. No
information is lost about the physical flow: if we know one point on a
trajectory, we can hop at will back and forth between the reduced
and the full \statesp\ trajectories, just as we can reconstruct a
continuous trajectory from its \PoincSec s.

At this point it is worth noting that imposing the global and fixed slice
\refeq{PCsectQ} is not the only way to separate equivariant dynamics into
`group dynamics' and `shape' dynamics\rf{BeTh04}. In modern mechanics
and even field theory (where elimination of group-directions is called
`gauge-fixing') it is natural to separate the flow {\em locally} into group dynamics
and a transverse, `horizontal' flow\rf{Smale70I,AbrMars78}, by the
`method of connections'\rf{rowley_reduction_2003}. From our point of
view, such approaches are not useful, as they do not reduce the dynamics
to a lower-dimensional \reducedsp\ $\pS/\Group$.

\section{\Sset}
	\label{sec:singul}

If  two patterns are close, their group orbits are nearly parallel, and
$\braket{\groupTan(\ssp)}{\sliceTan{}} \neq 0$. Hence a {\slice} is
transverse to all group orbits in an open neighborhood of the {\template}
\slicep, but not so {\em globally}. As we go away from
the {\template} point, the angles of the group orbit traversals
can decrease all the way to zero, until their group tangents lie in the slice.
This set of points defines a purely group-theoretic boundary of
the {\template's} neighborhood (every point has a group orbit, the dynamics
plays no role here, only the notion of distance).
Furthermore, whenever the group tangent of the \reducedsp\ trajectory
points into the slice, the denominator in \refeq{eq:so2reduced} vanishes
and {\angVel} $\dot{\gSpace}$ is not defined. We now show that these
singularities (a) also lie in a hyperplane, determined by the symmetry
group alone, and (b) induce computable jumps in the \reducedsp\
trajectory.

 													\toCB
Two hyperplanes sketched in \reffig{fig:slice}\,(b) are associated with
any given {\template} \slicep; the slice \refeq{PCsectQ}, and the
hyperplane of points \sspSing\ defined by being normal to  the quadratic
Casimir-weighted vector $\Lg^2\slicep$, such that from the {\template}
vantage point their group orbits are not transverse, but locally
`horizontal,'
\beq
\braket{\groupTan(\sspSing)}{\sliceTan{}}
 =
{-\braket{\sspSing}{\Lg^2\slicep}}
 =0
\ee{sliceSingl0}
(for simplicity, in this section we specialize to the  $\SOn{2}$ case).

We shall refer to the
intersection of the two as the
{\em \sset} $S$, \ie, the set of all points $\sspRSing$ which are both
{(a)} in the {\slice}, and {(b)} whose group tangent $\groupTan(\sspRSing)$
is also in the  {\slice}:
\bea
\braket{\sspRSing}{\sliceTan{}}&=&0 \continue
\braket{\groupTan(\sspRSing)}{\sliceTan{}}
 &=&
{-\braket{\sspRSing}{\Lg^2\slicep}}
 =0
\label{sliceSingl}
\eea
(this is called `singular set' in \refref{SiCvi10}).
Looking back at \refeq{SO2inflPoint}, we see that $S$ is the locus of
inflection points, a hyperplane through which the curvature of the
distance function changes sign, and a local minimum turns into a local
maximum.  For example, for the \cLe\ $\sspRSing =
(x_1^*,x_2^*,y_1^*,y_2^*,z^*)$, $\slicep = (x_1',x_2',y_1',y_2',z')$, and
the 3\dmn\ {\sset} $\sspRSing \in S \subset \pSRed$ is given by vanishing
denominator in \refeq{braketCL}:
\[
0 = {x_1^* x'_1+x_2^* x'_2+y_1^* y'_1+y_2^* y'_2}
\,.
\]
The {\sset}  $S$ is purely an artifact of the choice of a {\template},
and has nothing to do with the dynamics; whenever the full \statesp\
trajectory crosses an {\sset}, it pays it no heed whatsoever.

We next show that the singularity in
the formula \refeq{SL:CLEsliceRot} causes a discontinuous jump in the
moving frame angle $\gSpace$. Consider a full \statesp\ trajectory
$\ssp(\tau)$ that passes through the {\sset} \refeq{sliceSingl} at time $\tau^*$,
$\sspRSing =\ssp(\tau^*)$, which we shall set to  $\tau^*=0$. At that
instant the moving frame angle $\gSpace$ is formally undefined: the
numerator $\braket{\sspRSing}{\sliceTan{}}$ in \refeq{SL:CLEsliceRot}
vanishes ($\sspRSing$ is in the slice and satisfies the slice condition
\refeq{PCsectQ}), and the denominator
${\braket{\groupTan_{}(\sspRSing)}{\sliceTan{}}}$ vanishes by
\refeq{sliceSingl}. Nevertheless, the trajectory going through the singularity
is well defined, as in the linear approximation the numerator and the
denominator in the moving frame formula \refeq{SL:CLEsliceRot} are given
by
\bea
\braket{\ssp}{\sliceTan{}}
    &=&
\braket{(\sspRSing+\vel(\sspRSing) \, \tau )}
       {\sliceTan{a}}
    \,=\,  \braket{\vel(\sspRSing)}
       {\sliceTan{a}} \, \tau
	\label{singSetVelo}\\
\braket{\groupTan(\ssp)}{\sliceTan{}}
    &=&
\braket{\groupTan(\sspRSing+\vel(\sspRSing) \, \tau )}
       {\sliceTan{a}}
    \,=\, \braket{\groupTan(\vel(\sspRSing))}
       {\sliceTan{a}} \, \tau
\,.
\label{singSetSign}
\eea
In other words, the moving frame rotates the \statesp\ point $\ssp$ and
the velocity $\vel(\ssp)$ by the same angle, so either can be used to
compute it. The shortest distance condition demands that we pick the
solution with positive curvature \refeq{SO2inflPoint}, so
$\braket{\groupTan(\ssp)}{\sliceTan{}}\geq 0$
for all $\tau$. As the trajectory traverses $\sspRSing$, the time $\tau$
changes the sign from negative to positive; hence we must switch from the
solution $\sspRSing$ to another extremum for which
$\braket{\groupTan(\vel(\sspRSing))}{\sliceTan{}}$ is positive. In the
\cLf\ example there are only two extrema $\{\gSpace,\gSpace+\pi\}$,
so the moving frame angle $\gSpace$ jumps discontinuously by $\pi$.

Within the \reducedsp\  an {\sset} crossing has a dramatic effect: the
\reducedsp\ flow $\sspRed(\tau)$ \emph{jumps} whenever it crosses the
{\sset} $\sspRSing$, by an amount that we now compute.
Suppose that a \reducedsp\ trajectory passes through a singularity
$\sspRSing$ at time $\tau=\tau^*$. At
that instant the numerator $\braket{\vel(\sspRSing)}{\sliceTan{a}}$ is
(generically) finite, but as the group tangent of the point $\sspRSing$
lies in the slice, ${\braket{\groupTan_{}(\sspRSing)}{\sliceTan{}}}=0$
by \refeq{sliceSingl}, the denominator in \refeq{eq:so2reduced} vanishes,
and the {\angVel} $\dot{\gSpace}$ shoots off to infinity.

 \begin{figure}
 \begin{center}
(a) \includegraphics[width=0.35\textwidth]{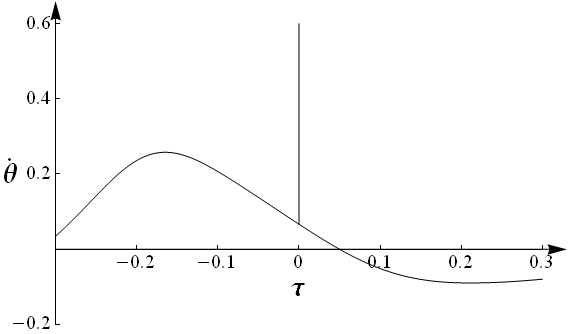}%
~~
(b) \includegraphics[width=0.35\textwidth]{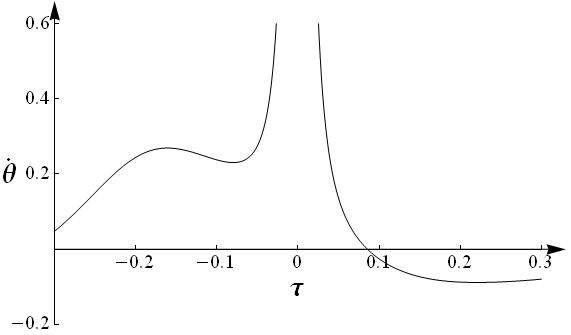}%
 \end{center}
 \caption{\label{fig:dthetasing}
The {\angVel} $\dot{\gSpace}$ for two \cLf\
\reducedsp\ trajectories in a slice defined by the
{\template} $\slicep$ given in \refeq{exmplTempl}:
 (a) As the trajectory $\sspRed(\tau)$ passes through the
singular point  $\sspRSing$ given in \refeq{exmplTempl},
the {\angVel} diverges
$\dot{\gSpace} \to \infty$ as a Dirac delta function.
(b) The {\angVel} for a nearby trajectory going
through $\sspRSing+\delta \ssp$,
$\delta\ssp=(0.01,0,0,0,0)$ exhibits a large
but finite excursion close to the singularity.
 }%
 \end{figure}

\begin{figure}
 \begin{center}
(a) \includegraphics[width=0.42\textwidth]{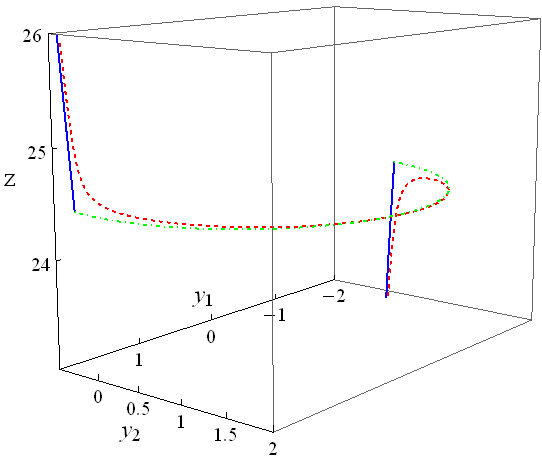}%
~~~~~~~~
(b)~ \includegraphics[width=0.40\textwidth]{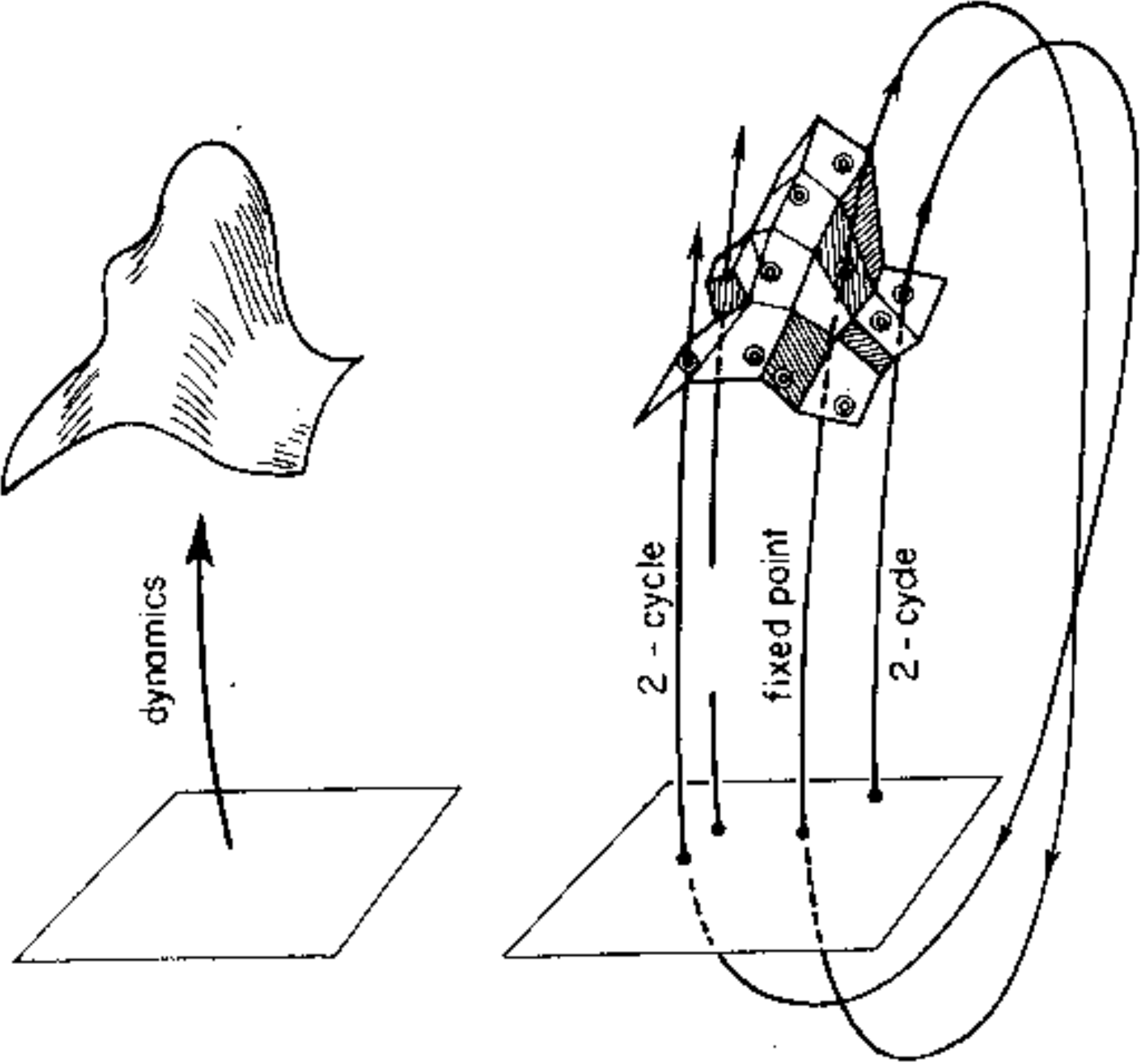}
 \end{center}
 \caption{\label{fig:singpass}
{(color online).}
(a)
Blow-up of a jump in \reffig{fig:Fullspace}\,(b), indicated by a small
rectangle.
{(blue/full line)}
A trajectory that passes through the singular point
$\sspRSing$ given in \refeq{exmplTempl}. Note the instantaneous jump in
the trajectory,  caused by the divergence in velocity
(\reffig{fig:dthetasing}\,(a)) as the trajectory traverses the {\sset}.
The neighboring red/dashed trajectory going through $\sspRSing+\delta
\ssp$, $\delta \ssp =(0,0.025,0,0,0)$, makes a rapid transit around the
singularity. The {green/dotted} trajectory is the group orbit of $\sspRSing$
between the two $\gSpace$ that rotate $v(\sspRSing)$ in the slice. Note
also how the red/dashed trajectory begins near the {blue/full}
trajectory, closely follows the {green/dotted} trajectory after the singularity
point, reaches the other side of the {green/dotted} arc and then resumes
closely following the {blue/full} trajectory.
(b)
Smooth dynamics  (left frame) tesselated by the skeleton of periodic
points, together with their linearized neighborhoods, (right frame).
Indicated are segments of two 1-cycles and a 2-cycle that alternates
between the neighborhoods of the two 1-cycles, shadowing first one, and
then the other
(from \wwwcb{}).
 }%
 \end{figure}

As an illustration of such jump, consider a blow-up of the small rectangle indicated
in \reducedsp\ flow \reffig{fig:Fullspace}\,(b). Here the {\template}
\bea
\slicep 	&=& (0.887846,-0.150461,0.4,-0.12,0)
	\continue
\sliceTan{} &=& (0.150461,0.887846,0.12,0.4,0)
	\label{exmplTempl} \\
\sspRSing	&=& (-0.889135, -0.0401956, 1.91332, -0.150327, 24.4436)
\nnu
\eea
was reverse-engineered, by picking a point $\sspRSing = \ssp(0)$ from a
segment of the full \statesp\ ergodic trajectory $\ssp(\tau)$ and then
computing \slicep\ such that $\sspRSing$ lies in the {\sset}.
    {
As the trajectory $\sspRed(\tau)$ passes through $\sspRSing$, the moving
frame $\gSpace$ jumps by $\pi$. Any trajectory nearby $\ssp(\tau)$ in the
full space (for example, the red/dashed trajectory in \reffig{fig:singpass}\,(a)) is
assigned a nearby $\gSpace$ both before and after $\sspRed(\tau)$ passes
through $\sspRSing$. The closer the trajectory is to $\ssp(\tau)$ in the
full space, the shorter the time interval where its moving frame differs
significantly from $\ssp(\tau)$'s.  If its symmetry-\-reduced trajectory does not
pass through inflection hyperplane $S$, then $\gSpace$ is continuous and
must change by $\pi$ in a short interval of time that shrinks the closer
the trajectory is to $\ssp(\tau)$. Hence,
    }
as the trajectory $\sspRed(\tau)$ passes through the $\sspRSing$, the {\angVel}
diverges $\dot{\gSpace} \to \infty$ as a Dirac delta function,
\reffig{fig:dthetasing}\,(a), and the \reducedsp\ trajectory goes through
the inflection \refeq{sliceSingl} and jumps to the $\pi$-rotated extremum
of the distance function, \reffig{fig:singpass}\,(a).

The {\sset} $S$ is the intersection of two hyperplanes: (1) the slice
(the shortest distance from the group-orbit to the {\template}), and (2)
the closest inflection in the distance function \refeq{sliceSingl}. While
all group orbits of a  generic trajectory cross the slice, the trajectory
has vanishing probability to cross the lower-dimensional {\sset} - that
is why we had to `engineer' the slice \refeq{exmplTempl}. However, an
ergodic trajectory  might come arbitrarily close to  $S$ arbitrarily
often. Such nearby \reducedsp\ trajectories exhibit large {\angVels}
$\dot{\gSpace}$, \reffig{fig:dthetasing}\,(b), and very fast, nearly
semi-circular excursions close to the singularity,
\reffig{fig:singpass}\,(a). Which segment of the group orbit they follow
depends on the side from which the trajectory approached the {\sset}.

In summary, whenever a \reducedsp\ trajectory crosses the {\sset}, it
jumps instantaneously and discontinuously to the new group orbit point
with the shortest distance to the template. We
have shown that these jumps are harmless and theoretically under control.
Nearby trajectories are numerically under control if sufficient care is
taken to deal with large \angVels. But they are an artifact of the
\mslices\ of no dynamical significance, and an uncalled-for numerical
nuisance. We now outline a strategy how to avoid them altogether by a
clever choice of a {\em set} of {\template s}.

\section{Charting the \reducedsp}
	\label{sec:chart}

So far, the good news is that for a generic {\template} $\slicep$ (\ie,
any $\slicep$ whose group orbit has the full $N$-dimensions of the
symmetry group \Group), the slice hyperplane \refeq{PCsectQ} cuts across
the group orbit of {\em every} point in the full \statesp\ \pS. But is
this a useful symmetry reduction of the full \statesp? A distant pattern
that is a bad match to a given {\template} will have any number of
locally `minimal' distances, each yet another bad match. Physically it
makes no sense to use a single slice (a set of all group orbit points
that are closest to one given {\template}) globally.

Work on \KS\ and the work of Rowley and
Marsden\rf{rowley_reconstruction_2000} suggests how to proceed: it was
shown in \refrefs{lanCvit07,SCD07} that for turbulent/chaotic systems a
set of Poincar\'e sections is needed to capture the dynamics. The choice
of sections should reflect the dynamically dominant patterns seen in the
solutions of nonlinear PDEs. We propose to construct a global atlas of
the symmetry \reducedsp\ $\pS/\Group$ by deploying both slices and
linear Poincar\'e sections across neighborhoods of the qualitatively most
important patterns, taking care that the {\template s} chosen have no
symmetry. Each slice $\pSRed{}^{(j)}$, tangential to one of a finite
number of {\template s}  $\slicep{}^{(j)}$, provides a local chart for a
neighborhood of an important, qualitatively distinct class of solutions
(2-rolls states, 3-rolls states, \etc); together they `Voronoi'
tessellate  the curved manifold in which the reduced strange attractor is
embedded by a finite set of hyperplane
tiles\rf{rowley_reconstruction_2000,RoSa00}. This is the symmetry-reduced
generalization of the idea of {\statesp\ tessellation} by a set of
periodic-orbits, so dear to a professional cyclist,
\reffig{fig:singpass}\,(b).

So how do we propose to implement this tessellation?

The physical task is to, for a given dynamical flow, pick a set of
qualitatively distinct {\template s} whose slices are locally tangent to
the strange attractor. A `slice' is a purely group-theoretic, linear
construct, with no reference to dynamics; a given {\template}
$\slicep{}^{(1)}$ defines the associated slice $\pSRed$, a
($d\!-\!1$)\dmn\ tangent hyperplane (for simplicity, in this section we
specialize to the $\SOn{2}$ case). Within it, there is a ($d\!-\!2$)\dmn\
{\sset} \refeq{sliceSingl}. If we pick another {\template} point
$\slicep{}^{(2)}$, it comes along with its own slice and {\sset}. Any
neighboring pair of $(d\!-\!1)$\dmn\ slices intersects in a `ridge'
(`boundary,' `edge'), a $(d\!-\!2)$\dmn\ hyperplane, easy to compute.
A global atlas so constructed should be sufficiently
fine-grained that we never hit any {\sset} singularities. The {\sset}s
should be eliminated by requiring that they lie either on the far sides
of the slice-slice intersections, or elsewhere where the strange
attractor does not tread. Each `chart' or `tile,' bounded by ridges to
neighboring slices, should be sufficiently small so that the {\sset} is
nowhere within the part of the slice explored by the strange attractor.

Follow an ant as it traces out a symmetry-reduced trajectory
$\sspRed{}^{(1)}(\tau)$, confined to the slice $\pSRed{}^{(1)}$. The
moment $\braket{\sspRed{}^{(1)}(\tau)}{\sliceTan{}{}^{(2)}}$ changes
sign, the ant has crossed the ridge, we symmetry-reduce with respect to
the second slice, and the ant continues its merry stroll within the
$\pSRed{}^{(2)}$ slice. Or, if you prefer to track the  given full
\statesp\ trajectory $\ssp(\tau)$, you compute the moving-frame angle
with respect to each (global) slice, and check to which tile does the
given group orbit belong.

What about the fixed-point subspace $\pS_\Group$ (see \refeq{dscr:InvPoints})?
Because of it, the action of \Group\ is globally neither free nor proper,
\etc. All intersections of slices, ridges and {\sset s} contain the
fixed-point subspace $\pS_\Group$. Should we worry?
    {
There are spurious singularities that are artifacts of a linear slice,
described by the associated {\sset}, and there are genuine, symmetry induced
singularities, such as the embedding of an invariant subspace in the full
\statesp\ (here $z$-axis). A {\sset} includes the invariant subspace and
cannot `cure' those. Indeed, we have tried to construct an example of a
two-slice chart, but for \cLe\ we have not been able to find a good one.
As the trajectory approaches the $z$-axis from various directions, we
have not found of a way to choose two slices such that it the group orbit
is tangent to one but not to the other. This is not a serious problem, as
the Poincar\'e section and the associated return map\rf{SiCvi10}
can be chosen to lie away from either {\sset}.
    }

The objective of the \mslices\ is to freeze\rf{BeTh04} all equivariant
coordinates; once frozen, they together with the  $\pS_\Group$
coordinates span the symmetry-\reducedsp.

There is a rub, though - you need to know how to pick the phases of
neighboring {\template s}. This is a reflection of the flaw inherent in use
of a slice hyperplane globally: a slice is derived from the Euclidean
notion of distance, but for nonlinear flows the distance has to be
measured curvilinearly, along unstable
manifolds\rf{Christiansen97,DasBuch}. We nevertheless have to stick with
tessellation by linearized tangent spaces, as curvilinear charts appear
computationally prohibitive. Perhaps a glance at
\reffig{fig:singpass}\,(b) helps visualize the problem; imagine that the
trajectories drawn are group orbits, and that the tiles belong to the
slices through {\template} points on these orbits. One could slide
{\template s} along their group orbits until the pairs of straight line
segments connecting neighboring {\template} points are minimized, but
that is not physical: one would like the dynamical trajectories to cross
ridges as continuously as possible. So how is one to pick the phases of
the {\template s}? The phase of the first {\template} is for free, but
the moving frame transformation \refeq{sspOrbit} is global, and can be
applied only once. The choice of the first template thus fixes all {\em
relative phases} to the succeeding {\template s}, as was demonstrated in
\refref{SCD07}: the universe of all other solutions is rigidly fixed
through a web of heteroclinic connections between them. This insight
garnered from study of a 1-dimensional \KS\ PDE is more remarkable still
when applied to the plane Couette flow\rf{GHCW07}, with 3-$d$ velocity
fields and two translational symmetries. The {\em relative phase} between
two {\template s} is thus fixed  by the shortest heteroclinic connection,
a rigid bridge from one neighborhood to the next. Once the relative phase
between the templates {\template s} is fixed, so are their their slices,
\ie, their tangent hyperplanes, and their intersection, \ie, the  ridge
joining them.

\section{What lies ahead} 
    \label{sec:concl}

Many physically important spatially-extended and fluid dynamics systems
exhibit continuous symmetries. For example,  excitable
media\rf{ZaZha70,Winfree73,Winfree1980,BaKnTu90,Barkley94}, \KS\
flow\rf{ku,siv,SCD07}, {\pCf}\rf{Visw07b,GHCW07,HGC08,GibsonMovies}, and
pipe flow\rf{Wk04,Kerswell05} are invariant (equivariant) under
combinations of translational (Euclidean), rotational and discrete
symmetries. If a physical problem has a symmetry, one should use it - one
does not want to compute the same solution over and over, all one needs
is to pick one representative solution per each symmetry related
equivalence class. Such procedure is called symmetry reduction.  In this
paper we have investigated symmetry reduction by the \mslices, a linear
procedure particularly simple and practical to implement, and answered
affirmatively the two main questions about the method:
(1) does a slice cut the group orbit of \emph{every} point in the
dynamical \statesp?
(2) can one deal with the {\sset s} that the method necessarily
introduces?

We have shown here that a symmetry-reduced trajectory passes through such
singularities through computable jumps, a nuisance numerically, but cause
to no conceptual difficulty. However, while a slice intersects each group
orbit in a neighborhood of a {\template} only once, extended globally any
slice intersects every group orbit multiple times. So even though every
slice cuts all group orbits, it makes no sense physically to use one
slice (a set of \emph{all} group orbit points that are closest to a given
{\template}) globally. We propose instead to construct a global atlas by
deploying sets of slices and linear Poincar\'e sections as charts of
neighborhoods of the most important (relative) equilibria and/or
(relative) periodic orbits.

Such global atlas should be sufficiently fine-grained so that an
unstable, ergodic trajectory never gets too close to any of the {\sset
s}. Why does this proposal have none of the elegance of, let's say,
Killing-Cartan classification of simple Lie algebras? Why is this
symmetry reduction purely a numerical procedure, rather than an analytic
change of equivariant coordinates to invariant ones? The theory of
\emph{linear} representations of compact Lie groups is a well developed
subject, but role of symmetries in \emph{nonlinear} settings is
altogether another story. It is natural to express a dynamical system
with a symmetry in the symmetry's linear eigenfunction basis (let  us
say, Fourier modes), but for a nonlinear flow different modes are
strongly coupled, and group orbits embedded in such coordinate bases can
be highly convoluted, in ways that no single global slice
hyperplane can handle intelligibly.

It should be emphasized that the atlas so constructed retains the
dimensionality of the original problem. The full dynamics is faithfully
retained, we are \emph{not} constructing a lower-dimensional model of the
dynamics. Neighborhoods of unstable \eqva\ and \po s are dominated by
their unstable and least contracting stable eigenvalues and are, for all
practical purposes, low-dimensional. Traversals of the ridges are,
however, higher dimensional. For example, crossing from the neighborhood
of a two-rolls state into the neighborhood of a three-rolls state entails
going through a pattern `defect,' a rapid transient whose precise
description requires many Fourier modes. Nevertheless, the recent
progress on separation of `physical' and `hyperbolically isolated'
covariant Lyapunov
vectors\rf{PoGiYaMa06,ginelli-2007-99,YaTaGiChRa08,TaGiCh09} gives us
hope that the proposed atlas could provide a systematic and controllable
framework for construction of lower-dimensional models of `turbulent'
dynamics of dissipative PDEs.

While it has been demonstrated in \refref{SiCvi10}  that the \mframes,
with a judicious choice of the {\template} and {\PoincSec}, works for a
system as simple as the \cLf, one still has to show that the method can
be implemented for a truly high-dimensional flow.
Siminos\rf{SiminosThesis} has used a modified \mframes\ to compute
analytically a 128\dmn\ invariant basis for \reducedsp\ $\pSRed
= \pS/\SOn{2}$, and shown that the unstable manifolds of \reqva\ play
surprisingly important role in organizing the geometry of \KS.
In \refref{SCD07} it
was found that the coexistence of four \eqva, two \reqva\
(traveling waves) and a
nested \fixedsp\ structure in an effectively $8$-dimensional \KS\ system
complicates matters sufficiently that no symmetry reduction by the \mslices\ has been
attempted so far.
More importantly, a symmetry reduction of pipe flows, which
due to the translational symmetry have only relative (traveling)
solutions, remains an outstanding challenge\rf{ACHKW11}.

	\medskip
	\noindent{\bf Acknowledgments}
We sought in vain Phil Morrison's sage counsel on how to reduce
symmetries, but none was forthcoming - hence this article. We are
grateful to
D.~Barkley,
W.-J.~Beyn,
C.~Chandre,
K.A.~Mitchell,
B.~Sandstede,
R.~Wilczak,
and in particular E.~Siminos and R.L.~Davidchack
for spirited exchanges.
S.F. work was supported by the National Science Foundation grant
DMR~0820054 and a Georgia Tech President's Undergraduate Research Award.
P.C. thanks Glen Robinson Jr. for support. 	
\appendix
\section{Symmetries of dynamics}
	\label{sec:SymmDyn}

In this Appendix we review a few basic facts about dynamics and
symmetries. We follow notational conventions of
Chaosbook.org\rf{DasBuch}, to which the reader is referred to for a more
extensive discussion of dynamics and symmetries.

If a pipe is rotated around its axis or translated, the shifted and
rotated state of the fluid is a physically equivalent solution of
the Navier-Stokes equations. Such rotations and translations
are examples of continuous symmetries. On the level of equations of
motion, one says that a flow $\dot{x}= \vel(x)$ is \emph{equivariant}
under a coordinate transformation $\LieEl$ if
\beq
\vel(x)=\LieEl^{-1}\vel(\LieEl \, x)
\,.
\ee{eq:FiniteRot}
The totality of elements
$\LieEl$ forms \Group, the {\em symmetry group} of the flow.
An element of a compact Lie group $\Group \subset \On{d}$ that is
continuously connected to the identity can be parametrized as
\beq
\LieEl(\gSpace)=e^{{\gSpace} \cdot \Lg }
    \,,\qquad
\gSpace \cdot \Lg = \sum_{a=1}^N \gSpace_a \Lg_a
\,,
\ee{FiniteRot}
where $\gSpace \cdot \Lg $ is a \emph{Lie algebra} element, $\gSpace =
(\gSpace_1,\gSpace_2,\cdots\gSpace_N)$ are the parameters (`phases,'
`angles,' `shifts') of the transformation, and the $\Lg_a$ are a set of
$N$ linearly independent $[d\!\times\!d]$ antisymmetric matrices acting
linearly on the {\statesp} vectors. A spatial transformation induced by
infinitesimal variations of group phases
$
\LieEl(\delta \gSpace) \simeq 1 + \delta \gSpace \cdot \Lg
\,.
$ 
is
\beq
\delta {\ssp} = \delta \gSpace \cdot \groupTan(\ssp)
\,,
\ee{PC:groupTan0}
where the $N$ vectors
\beq
 \groupTan_{a}(\ssp) = \Lg _{a} \ssp
    \,,\qquad
 a=1,2,\cdots,N,
\ee{PC:groupTan}
span the group tangent space at $\ssp$. We use $\groupTan_a(\ssp)$
notation (rather than $\Lg_{a}\ssp$) to emphasize that the group action
induces a \emph{tangent field} at $\ssp$.
A transformation induced by infinitesimal
time-dependent variations \refeq{PC:groupTan0} of group phases
can be thought of as an `{\angVel},'
$\delta \gSpace_a = \timeStep \, \dot{\gSpace_a}$ is
\beq
\dot{\ssp} = \dot{\gSpace} \cdot \groupTan(\ssp)
\,.
\ee{PC:groupTan1}

The {tangent field} is of dimension $N$, as long as the point $\ssp$ does
not belong to a fixed-point subspace. Points in the \emph{fixed-point
subspace}  $\pS_\Group$ are fixed points of the full group action. They
are called \emph{invariant points},
\beq
\pS_\Group = \Fix{\Group} =
   \{ \ssp \in \pS : {g} \, \ssp = \ssp \mbox{ for all } g \in \Group \}
\,,
\ee{dscr:InvPoints}
or, infinitesimally,  $\Lg_{a}\ssp=0$
{for $a=1,2,\cdots,N$}.
If a point is an invariant point
of the symmetry group, by equivariance the velocity at that point is also
in $\pS_\Group$, so the trajectory through that point will remain in
$\pS_\Group$. $\pS_\Group$ is disjoint from the rest of the {\statesp}
since no trajectory can ever enter or leave it.

Any representation of a compact group $\Group$ is
{fully reducible\rf{Hall03}}.
The invariant tensors constructed by contractions of $\Lg_a$ are useful
in identifying irreducible representations. The simplest such invariant
is
\beq
{\Lg} \cdot \Lg = - \sum_m C_2^{(m)} \, \id^{(m)}
\,,
\ee{QuadCasimir}
where $C_2^{(m)}$ is the quadratic Casimir for irreducible representation
labeled $m$, and $\id^{(m)}$ is the identity on the irreducible
subspace $m$, 0 elsewhere. For compact groups $C_2^{(m)}$ are strictly
nonnegative. $C_2^{(m)} =0$ if $m$ is an invariant subspace.

    %
    %
The simplest example of a Lie group is given by the action of \SOn{2} on
a smooth function $u(\gSpace + 2\pi) = u(\gSpace)$ periodic on interval
$[-\pi,\pi]$. Expand $u$ as a Fourier series
\beq
u(\gSpace) = \frac{a_0}{2} + \sum_{m=1}^\infty \left(
a_m \cos m \gSpace + b_m \sin m \gSpace
                               \right)
\,.
\ee{FourierExp}
The matrix representation of the \SOn{2}\ action
$\LieEl(\gSpace') u(\gSpace) = u(\gSpace+\gSpace')$
on the Fourier coefficient pair
$(a_m,b_m)$ is
	\ifarticle  
\bea
\LieEl^{(m)}(\gSpace')
    &=& \exp{\left({\gSpace'} \cdot \Lg^{(m)}\right)}
	\,=\,
   \left(\barr{cc}
 ~\cos m \gSpace'  & \sin m \gSpace' \\
 -\sin m \gSpace'  & \cos m \gSpace'
    \earr\right)
\continue
&=&
 \cos m \gSpace' \id^{(m)}
  + \sin m \gSpace'\, \frac{1}{m} \Lg^{(m)}
\,.
\label{SO2irrepAlg-m}
\eea
    \else  
\bea
\LieEl^{(m)}(\gSpace')
    &=& \exp{\left({\gSpace'} \cdot \Lg^{(m)}\right)}
	\,=\,
   \left(\barr{cc}
 ~\cos m \gSpace'  & \sin m \gSpace' \\
 -\sin m \gSpace'  & \cos m \gSpace'
    \earr\right)
\,=\,
 \cos m \gSpace' \id^{(m)}
  + \sin m \gSpace'\, \frac{1}{m} \Lg^{(m)}
\,.
\label{SO2irrepAlg-m}
\eea
	\fi
Here
\beq
 \Lg^{(m)} =   \left(\barr{cc}
    0  &  m  \\
   -m  &  0
    \earr\right)
\label{SO2irrepAlg-Lg}
\eeq
is the Lie algebra generator and $\id^{(m)}$ is the identity
on the irreducible subspace labeled $m$, 0 elsewhere. The \SOn{2}\ group
tangent $\groupTan(u)$ to \statesp\ point $u$ is
\beq
 \groupTan(u) = \sum_{m=1}^\infty \groupTan^{(m)}(u)
    \,,\qquad
 \groupTan^{(m)}(u)
\,=\, m \,\left(\barr{c}
   ~b_m  \\
   -a_m
    \earr\right)
\,,
\ee{u:x:tang}
{                                                  \toCB
and the quadratic Casimir \refeq{QuadCasimir} for irreducible representation
labeled $m$ is $C_2^{(m)} = m^2$.
    }

\section{Singularities of $\SOn{2} \times \SOn{2}$}
	\label{sec:singulProd}

Two groups \Group\ and $H$ can be combined into the {product group}
$\Group \times H$, whose elements are pairs $(\LieEl,h)$, where $\LieEl$
belongs to \Group, and $h$ belongs to $H$, with the group multiplication
rule
\(
(\LieEl_1,h_1)(\LieEl_2,h_2)=(\LieEl_1 \LieEl_2,h_1 h_2)
\,.
\)
Some important fluid-dynamical flows exhibit continuous symmetries which
are the products of $\SOn{2}$ groups, each of which acts on a subset  of
the {\statesp} coordinates. The \KS\ equations\rf{ku,siv},
{\pCf}\rf{Visw07b,GHCW07,HGC08}, and pipe
flow\rf{Wk04,Kerswell05} all have continuous symmetries of this form.

Let $\Group \times H$ be a Lie group with two sets of infinitesimal
generators, $\Lg_1$ and $\Lg_2$, such that the $\Lg_1$ acts only on the
$(a)$ coordinates ($e^{\gSpace_1 \Lg_1} \,
(a,b)=(e^{\gSpace_1\Lg_1}a,b)$), or, infinitesimally,
$\Lg_1(a,b)=(\Lg_1 a,0)$, and $\Lg_2$ acts only on the $(b)$ coordinates,
$\Lg_2(a,b)=(0,\Lg_2 b)$. Taken together, $\Lg_1 \Lg_2(a,b) = \Lg_2
\Lg_1(a,b) = (0,0) $ for all $(a,b)$, so $\Lg_1 \Lg_2=0$.

For simplicity, we now specialize to the  $\SOn{2} \times \SOn{2}$ case.
Suppose we are rotating a trajectory $\ssp(\tau)$ into the slice normal
to the group tangents at $\slicep$.
{Using the restrictions the slice imposes on the {\angVel}
\refeq{eq:slicecondition} yields}
$\braket{\vel(\sspRed)}{\sliceTan{1}}-\dot{\gSpace_1} \braket{
\groupTan_1(\sspRed)}{\sliceTan{1}}-\dot{\gSpace_2} \braket{
\groupTan_2(\sspRed)}{\sliceTan{1}}=0$. We have
$\braket{\groupTan_2(\sspRed)}{\sliceTan{1}}=0$ since $\Lg_1 \Lg_2=0$,
leaving us with the equation for $\dot{\gSpace_1}$,
\beq
\dot{\gSpace_1}=     {\braket{\vel(\sspRed)}{\sliceTan{1}}} /
                     {\braket{\groupTan_1(\sspRed)}{\sliceTan{1}}}
\,,
\eeq
and similarly for $\dot{\gSpace_2}$, the same as \refeq{eq:so2reduced}
for the rotation group consisting of only the rotations generated by
either $\Lg_1$ or $\Lg_2$.  This means a point being singular depends
only on whether or not it is singular in either of the slices normal to
only one of the group tangents, breaking up the problem of determining if
a point is singular into the same problem for each of the $\SOn{2}$
separately. In \refsect{sec:singul} we described what happens to the
\reducedsp\ trajectory as it passes through singularity of a single
$\SOn{2}$ symmetry group. Using this result we can handle the
singularities for the product of arbitrarily many $\SOn{2}$.

\bibliographystyle{elsarticle-num}  
\bibliography{../../bibtex/siminos}

\begin{thebibliography}{10}
\expandafter\ifx\csname url\endcsname\relax
  \def\url#1{\texttt{#1}}\fi
\expandafter\ifx\csname urlprefix\endcsname\relax\def\urlprefix{URL }\fi
\expandafter\ifx\csname href\endcsname\relax
  \def\href#1#2{#2} \def\path#1{#1}\fi

\bibitem{MorrGree80}
P.~J. Morrison, J.~M. Greene, Noncanonical {Hamiltonian} density formulation of
  hydrodynamics and ideal magnetohydrodynamics, Phys. Rev. Lett. 45 (1980)
  790--794, see also Phys. Rev. Lett. 48, 569 (1982).

\bibitem{Low58}
F.~E. Low, A {{Lagrangian}} formulation of the {{Boltzmann-Vlasov}} equation
  for plasmas, Proc. R. Soc. London A 248~(1253) (1958) 282--287.

\bibitem{CHHM98}
H.~Cendra, D.~D. Holm, M.~J.~W. Hoyle, J.~E. Marsden, The {Maxwell-Vlasov}
  equations in {Euler-Poincar\'e} form, J. of Math. Phys. 39 (1998) 3138--3157.

\bibitem{CBcontinuous}
P.~Cvitanovi\'{c}, Chapter ``{Relativity} for cyclists'', in \refref{DasBuch}
  (2011).

\bibitem{SiCvi10}
E.~Siminos, P.~Cvitanovi{\'c}, Continuous symmetry reduction and return maps
  for high-dimensional flows, Physica D 240 (2011) 187--198.

\bibitem{rowley_reduction_2003}
C.~W. Rowley, I.~G. Kevrekidis, J.~E. Marsden, K.~Lust, Reduction and
  reconstruction for self-similar dynamical systems, Nonlinearity 16 (2003)
  1257--1275.

\bibitem{SiminosThesis}
E.~Siminos, Recurrent spatio-temporal structures in presence of continuous
  symmetries, Ph.D. thesis, School of Physics, Georgia Inst. of Technology,
  Atlanta, \\\wwwcb{/projects/theses.html} (2009).

\bibitem{Wilczak09}
R.~Wilczak, Reducing the state-space of the complex {L}orenz flow, {NSF REU}
  summer 2009 project, {U. of Chicago}, \\ \wwwcb{/projects/Wilczak/blog.pdf}
  (2009).

\bibitem{GibMcCLE82}
J.~D. Gibbon, M.~J. McGuinness, The real and complex {Lorenz} equations in
  rotating fluids and lasers, Physica D 5 (1982) 108--122.

\bibitem{rowley_reconstruction_2000}
C.~W. Rowley, J.~E. Marsden, Reconstruction equations and the
  {Karhunen-Lo\'eve} expansion for systems with symmetry, Physica D 142 (2000)
  1--19.

\bibitem{ahuja_template-based_2007}
S.~Ahuja, I.~Kevrekidis, C.~Rowley, Template-based stabilization of relative
  equilibria in systems with continuous symmetry, J. Nonlin. Sci. 17 (2007)
  109--143.

\bibitem{FelsOlver98}
M.~Fels, P.~J. Olver, Moving coframes: {I. A} practical algorithm, Acta Appl.
  Math. 51 (1998) 161--213.

\bibitem{FelsOlver99}
M.~Fels, P.~J. Olver, Moving coframes: {II. R}egularization and theoretical
  foundations, Acta Appl. Math. 55 (1999) 127--208.

\bibitem{OlverInv}
P.~J. Olver, Classical Invariant Theory, Cambridge Univ. Press, Cambridge,
  1999.

\bibitem{CartanMF}
E.~Cartan, La m\'ethode du rep\`ere mobile, la th\'eorie des groupes continus,
  et les espaces g\'en\'eralis\'es, Vol.~5 of {Expos\'es} de {G\'eom\'etrie},
  Hermann, Paris, 1935.

\bibitem{Barkley94}
D.~Barkley, Euclidean symmetry and the dynamics of rotating spiral waves, Phys.
  Rev. Lett. 72 (1994) 164--167.

\bibitem{SCD07}
P.~Cvitanovi{\'c}, R.~L. Davidchack, E.~Siminos, On the state space geometry of
  the {Kuramoto-Sivashinsky} flow in a periodic domain, SIAM J. Appl. Dyn.
  Syst. 9 (2010) 1--33, \arXiv{0709.2944}.

\bibitem{HGC08}
J.~F. Gibson, J.~Halcrow, P.~Cvitanovi{\'c}, Equilibrium and traveling-wave
  solutions of plane {Couette} flow, J. Fluid Mech. 638 (2009) 243--266,
  \arXiv{0808.3375}.

\bibitem{DasBuchMirror}
P.~Cvitanovi{\'c}, Chapter ``{World} in a mirror'', in \refref{DasBuch} (2011).

\bibitem{FiSaScWu96}
B.~Fiedler, B.~Sandstede, A.~Scheel, C.~Wulff, Bifurcation from relative
  equilibria of noncompact group actions: {Skew} products, meanders, and
  drifts, Doc. Math. 141 (1996) 479--505.

\bibitem{StGo09}
M.~Stone, P.~Goldbart, Mathematics for Physics: A Guided Tour for Graduate
  Students, Cambridge Univ. Press, Cambridge, 2009.

\bibitem{Marsd92}
J.~E. Marsden, Lectures on Mechanics, Cambridge Univ. Press, Cambridge, 1992.

\bibitem{MarsdRat94}
J.~E. Marsden, T.~S. Ratiu, Introduction to Mechanics and Symmetry, Springer,
  New York, 1994.

\bibitem{BeTh04}
W.-J. Beyn, V.~Th\"ummler, Freezing solutions of equivariant evolution
  equations, SIAM J. Appl. Dyn. Syst. 3 (2004) 85--116.

\bibitem{Smale70I}
S.~Smale, Topology and mechanics, {I}., Inv. Math. 10 (1970) 305--331.

\bibitem{AbrMars78}
R.~Abraham, J.~E. Marsden, Foundations of Mechanics, Benjamin-Cummings,
  Reading, Mass., 1978.

\bibitem{lanCvit07}
Y.~Lan, P.~Cvitanovi{\'c}, Unstable recurrent patterns in
  {Kuramoto-Sivashinsky} dynamics, Phys. Rev. E 78 (2008) 026208,
  \arXiv{0804.2474}.

\bibitem{RoSa00}
S.~T. Roweis, L.~K. Saul, Nonlinear dimensionality reduction by locally linear
  embedding, Science 290 (2000) 2323--2326.

\bibitem{Christiansen97}
F.~Christiansen, P.~Cvitanovi\'{c}, V.~Putkaradze, Spatiotemporal chaos in
  terms of unstable recurrent patterns, Nonlinearity 10 (1997) 55--70,
  \arXiv{chao-dyn/9606016}.

\bibitem{DasBuch}
P.~Cvitanovi\'{c}, R.~Artuso, R.~Mainieri, G.~Tanner, G.~Vattay, Chaos:
  Classical and Quantum, Niels Bohr Inst., Copenhagen, 2011, {\wwwcb{}}.

\bibitem{GHCW07}
J.~F. Gibson, J.~Halcrow, P.~Cvitanovi{\'c}, Visualizing the geometry of
  state-space in plane {Couette} flow, J Fluid Mech. 611 (2008) 107--130,
  \arXiv{0705.3957}.

\bibitem{ZaZha70}
A.~N. Zaikin, A.~M. Zhabotinsky, Concentration wave propagation in
  2-dimensional liquid-phase self-oscillating system, Nature 225 (1970)
  535--537.

\bibitem{Winfree73}
A.~T. Winfree, Scroll-shaped waves of chemical activity in 3 dimensions,
  Science 181 (1973) 937--939.

\bibitem{Winfree1980}
A.~T. Winfree, The Geometry of Biological Time, Springer, New York, 1980.

\bibitem{BaKnTu90}
D.~Barkley, M.~Kness, L.~S. Tuckerman, Spiral wave dynamics in a simple model
  of excitable media: {T}ransition from simple to compound rotation, Phys. Rev.
  A 42 (1990) 2489--2492.

\bibitem{ku}
Y.~Kuramoto, T.~Tsuzuki, Persistent propagation of concentration waves in
  dissipative media far from thermal equilibrium, Progr. Theor. Phys. 55 (1976)
  365--369.

\bibitem{siv}
G.~I. Sivashinsky, Nonlinear analysis of hydrodynamical instability in laminar
  flames - {I}. {D}erivation of basic equations, Acta Astronaut. 4 (1977)
  1177--1206.

\bibitem{Visw07b}
D.~Viswanath, Recurrent motions within plane {Couette} turbulence, J. Fluid
  Mech. 580 (2007) 339--358, \arXiv{physics/0604062}.

\bibitem{GibsonMovies}
J.~F. Gibson, P.~Cvitanovi\'c, Movies of plane {Couette}, Tech. rep., Georgia
  Inst. of Technology, {\wwwcb{/tutorials}} (2011).

\bibitem{Wk04}
H.~Wedin, R.~R. Kerswell, Exact coherent structures in pipe flow, J. Fluid
  Mech. 508 (2004) 333--371.

\bibitem{Kerswell05}
R.~R. Kerswell, Recent progress in understanding the transition to turbulence
  in a pipe, Nonlinearity 18 (2005) R17--R44.

\bibitem{PoGiYaMa06}
A.~Politi, F.~Ginelli, S.~Yanchuk, Y.~Maistrenko, From synchronization to
  {L}yapunov exponents and back, Physica D 224 (2006) 90, \arXiv{nlin/0605012}.

\bibitem{ginelli-2007-99}
F.~Ginelli, P.~Poggi, A.~Turchi, H.~Chat\'{e}, R.~Livi, A.~Politi,
  Characterizing dynamics with covariant {L}yapunov vectors, Phys. Rev. Lett.
  99 (2007) 130601, \arXiv{0706.0510}.

\bibitem{YaTaGiChRa08}
H.-l. Yang, K.~A. Takeuchi, F.~Ginelli, H.~Chat\'{e}, G.~Radons, Hyperbolicity
  and the effective dimension of spatially-extended dissipative systems, Phys.
  Rev. Lett. 102 (2009) 074102, \arXiv{0807.5073}.

\bibitem{TaGiCh09}
K.~A. Takeuchi, F.~Ginelli, H.~Chat\'{e}, Lyapunov analysis captures the
  collective dynamics of large chaotic systems, Phys. Rev. Lett. 103 (2009)
  154103, \arXiv{0907.4298}.

\bibitem{ACHKW11}
A.~P. Willis, A.~Avila, P.~Cvitanovi{\'c}, B.~Hof, Reduction of continuous
  symmetries of pipe flows by the method of slices, in preparation (2011).

\bibitem{Hall03}
B.~C. Hall, Lie Groups, Lie Algebras, and Representations, Springer, New York,
  2003.

\end{thebibliography}

\end{document}